\documentstyle[aps,prb,floats,epsf]{revtex}

\begin{document}
\draft


\twocolumn[\hsize\textwidth\columnwidth\hsize\csname@twocolumnfalse%
\endcsname
\title{Impurity-induced moments in underdoped cuprates}
\author{G.\ Khaliullin}
\address{Max-Planck-Institut f\"ur Physik komplexer Systeme, 
N\"othnitzer Str.\ 38, D-01187 Dresden, Germany\\
and Kazan Physicotechnical Institute of the Russian
Academy of Sciences, 420029 Kazan, Russia}
\author{R.\ Kilian}
\address{Max-Planck-Institut f\"ur Physik komplexer Systeme, 
N\"othnitzer Str.\ 38, D-01187 Dresden, Germany}
\author{S.\ Krivenko}
\address{Kazan Physicotechnical Institute of the Russian
Academy of Sciences, 420029 Kazan, Russia}
\author{P.\ Fulde}
\address{Max-Planck-Institut f\"ur Physik komplexer Systeme, 
N\"othnitzer Str.\ 38, D-01187 Dresden, Germany}
\date{02 July 1997}
\maketitle


\begin{abstract}
We examine the effect of a nonmagnetic impurity in a two-dimensional
spin liquid in the spin-gap phase, employing a drone-fermion 
representation of spin-$1/2$ operators. The properties
of the local moment induced in the vicinity of the impurity are
investigated and an expression for the 
nuclear-magnetic-resonance Knight shift is derived,
which we compare with experimental results. 
Introducing a second impurity into the spin liquid an
antiferromagnetic interaction between the moments is found
when the two impurities are located on different sublattices.
The presence of many impurities leads to a screening of this 
interaction as is shown by means of a coherent-potential 
approximation. Further, the Kondo screening of an impurity-induced
local spin by charge carriers is discussed.
\end{abstract}

]


\section{Introduction}
%
%
Substitution of Cu ions in the conduction planes of high-$T_c$ cuprates
by different nonmagnetic ions
presents an important experimental tool for the study of the
metallic state of these strongly correlated systems.
Unusual effects have been revealed especially when
these materials were doped with Zn, Al, or Ga. 
Among these interesting features is the appearance of a magnetic moment 
derived from the observation of a Curie-Weiss behavior of the
magnetic susceptibility. \cite{XIA87} Further studies
have been performed by electron paramagnetic resonance \cite{FIN90} 
(EPR) and nuclear magnetic resonance \cite{ALL91,ZHE93,MAH94,ISH96}
(NMR) experiments. Measurements of the NMR Knight shift indicate
that the impurity-induced local moments reside predominantly on Cu
sites neighboring the dopant. \cite{MAH94,ISH96}
It has been argued that the appearance of these local moments
can account for a significant reduction of $T_c$. \cite{FIN90}

On the theoretical side, impurity-induced moments 
have been studied for a variety of quantum disordered spin systems, 
including, for instance, spin ladders and spin-Peierls
systems, \cite{SIG96,NAG96} as well as underdoped 
cuprates. \cite{NAG95,KRI95b}
A common feature of these systems is the existence of a (pseudo-) 
gap in the spin-excitation spectrum. This spin gap is related
to the singlet-pairing correlations in the ground state, which
can be described in terms of resonance valence bonds \cite{AND87} (RVB)
or a valence-bond solid (VBS) state. \cite{AFF87}
Within this picture
the appearance of a Curie-type susceptibility can be
understood as an unpaired spin that is generated 
by the substitution of a spin site by a vacancy.

In the present paper we investigate the effect of nonmagnetic 
impurities on the local magnetic properties of weakly doped
cuprates. In order to describe the spin degrees of freedom of these 
systems we start with the Heisenberg antiferromagnet on a square lattice,
employing a drone-fermion representation of
spin-$1/2$ operators. \cite{MAT88} 
A mean-field decoupling of the Hamiltonian leads
to solutions corresponding to a RVB state in the flux 
phase \cite{AFF88,KOT88} that exhibits a pseudogap in the 
spin-excitation spectrum. A single spin vacancy introduced
into the spin liquid produces a local moment which is 
predominantly located
on the sites close to the vacancy. We derive
an expression for the impurity-induced NMR Knight shift
and fit it to experimental data. Introducing a second vacancy into 
the spin liquid leads to an antiferromagnetic interaction 
between the two moments. In
the presence of many impurities this interaction is screened
as is shown by using a coherent-potential approximation (CPA).
We further analyze the effect of nonmagnetic impurities on the 
relaxation of the nuclear magnetic moment in
terms of local spin fluctuations induced by the vacancies.
Finally, we discuss qualitatively the low-temperature
Kondo screening of the impurity-induced spins by charge carriers.
\nopagebreak

\section{The Model}
%
%
We start from the Hamiltonian of a two-dimensional
square lattice spin-$1/2$ Heisenberg antiferromagnet
\begin{equation}
\label{Heis}
H = J \sum_{\langle ij \rangle} \bbox{S}_i \cdot \bbox{S}_j ,
\end{equation}
where $\langle\,\,\rangle$ indicates summation over pairs of nearest-neighbor sites. 
The spin operators in the Hamiltonian (\ref{Heis}) are commonly
expressed in terms of pseudo-fermions
which requires us to impose the restriction that each site is occupied
by exactly one pseudoparticle.
Since it is rather difficult to account for these local 
constraints one usually treats them only on the average.
Here we will follow a different approach \cite{KRI95a} by employing
a drone-fermion description for spin-$1/2$ operators given 
by \cite{MAT88}
\begin{equation}
\label{Drone}
S_i^+ = f_i^{\dagger} \chi_i, \quad
S_i^- = \chi_i f_i, \quad
S_i^z = f_i^{\dagger} f_i - \textstyle{\frac{1}{2}},
\end{equation}
where $S_i^{\pm} = (S_i^x \pm iS_i^y)/\sqrt{2}$. The operator 
$f_i^{\dagger}$ creates a spinless fermion at site $i$;
the presence (absence) of an {\it f}-fermion corresponds to an up (down)
spin state. The real drone-fermion operator $\chi_i = \chi_i^{\dagger}$ 
with commutation rules $[\chi_i,\chi_j]_+ = \delta_{ij}$ is needed to
provide the proper commutation rules for spin operators on different sites.
Expressing the Hamiltonian (\ref{Heis}) in terms of the representation 
(\ref{Drone}) one obtains
\begin{eqnarray}
H &=& J \sum_{\langle ij \rangle} \Big\{ \left( 
f_i^{\dagger} \chi_i \chi_j f_j + 
\mbox{H.c.} \right) \nonumber\\
&&\mbox{}+ \left( f_i^{\dagger} f_i - {\textstyle \frac{1}{2}} \right)
\left( f_j^{\dagger} f_j - {\textstyle \frac{1}{2}} \right) \Big\}.
\label{Heis2}
\end{eqnarray}

Mean-field solutions to the Hamiltonian (\ref{Heis2}) corresponding
to the RVB states of conventional fermion mean-field theory
are found by employing a Hartree-Fock factorization 
with bond parameters
$i\Delta_{ij} = \langle\chi_j\chi_i\rangle$ and 
$i\tilde{\Delta}_{ij} = \langle f_j^{\dagger} f_i\rangle$
defined for nearest-neighbor pairs, yielding the 
following mean-field Hamiltonian: \cite{KRI95a}
\begin{eqnarray}
H_{\mbox{\scriptsize MF}} &=& 
J \sum_{\langle ij \rangle} \Big\{ \Big( -i \big( 
\tilde{\Delta}_{ij} + 
\Delta_{ij} \big) f_i^{\dagger} f_j \nonumber\\
&&\mbox{}- i\tilde{\Delta}_{ij} \chi_i\chi_j + \mbox{H.c.} \Big) 
+ 2\tilde{\Delta}_{ij} \Delta_{ij} + 
\tilde{\Delta}_{ij}^2 \Big\}.
\label{HMF}
\end{eqnarray}
For spin rotational symmetry of the ground state 
$\tilde{\Delta}_{ij} = \Delta_{ij}$ holds. 
In the following we assume the order parameter to 
depend only on relative site indices: $\Delta_{ij} =
\Delta_{\bbox{\delta}}$ with 
$\bbox{\delta} \in \{\hat{\bbox{x}},-\hat{\bbox{x}},
\hat{\bbox{y}},-\hat{\bbox{y}}\}$, where $\hat{\bbox{x}}$ and
$\hat{\bbox{y}}$ are the lattice unit vectors.

Dividing the square lattice into two sublattices {\it A} and {\it B}, we
define $i \in A$ and $j \in B$. The mean-field Hamiltonian
(\ref{HMF}) can then be diagonalized by expressing the 
$f$ and $\chi$ operators in momentum representation and employing 
the canonical transformation
\begin{eqnarray}
f_{\bbox{k}}^A &=& \left( a_{\bbox{k}} + b_{\bbox{k}} \right) / 
\sqrt{2}, \nonumber\\
\label{trans3}
f_{\bbox{k}}^B &=& \left( a_{\bbox{k}} - b_{\bbox{k}} \right) 
\exp(i\varphi_{\bbox{k}}) / \sqrt{2}, \\
\chi_{\bbox{k}}^A &=& \left( c_{\bbox{k}} + d_{\bbox{k}}\right) / 
\sqrt{2}, \nonumber\\
\label{trans4}
\chi_{\bbox{k}}^B &=& \left( c_{\bbox{k}} - d_{\bbox{k}}\right) 
\exp(i\varphi_{\bbox{k}}) / \sqrt{2},
\end{eqnarray}
where $\exp(i\varphi_{\bbox{k}}) = 
\Delta_{\bbox{k}}^*/|\Delta_{\bbox{k}}|$
with the definition
$\Delta_{\bbox{k}} = i \sum_{\bbox{\delta}} \Delta_{\bbox{\delta}} 
\exp(-i\bbox{k}
\bbox{\delta})$. Thus, the Hamiltonian (\ref{HMF}) becomes
\begin{eqnarray}
H_{\mbox{\scriptsize MF}} &=& \sum_{\bbox{k}\in \mbox{\scriptsize MBZ}} 
\xi_{\bbox{k}} \left( -a_{\bbox{k}}^{\dagger} a_{\bbox{k}} + 
b_{\bbox{k}}^{\dagger} b_{\bbox{k}} 
\right) \nonumber \\
&&\mbox{}+ \sum_{\bbox{k} \in \mbox{\scriptsize MBZ/2}} \xi_{\bbox{k}} 
\left( -c_{\bbox{k}}^{\dagger} 
c_{\bbox{k}} + d_{\bbox{k}}^{\dagger} d_{\bbox{k}} \right)
\label{Dia}
\end{eqnarray}
with $\xi_{\bbox{k}} = 2J|\Delta_{\bbox{k}}|$. The $k$ summations
extend over the full and half magnetic Brillouin zone, respectively, and
the constant terms in Eq.\ (\ref{HMF}) have been omitted.

The fermionic spectrum $E(\bbox{k}) = \pm \xi_{\bbox{k}}$ is determined
by the symmetry exhibited by the bond parameter 
$\Delta_{\bbox{\delta}}$. In a phase of mixed symmetry defined by
\begin{equation}
\Delta := \Delta_{\hat{\bbox{x}}} = -\Delta_{-\hat{\bbox{x}}} = 
\Delta_{\hat{\bbox{y}}} = \Delta_{-\hat{\bbox{y}}}
\end{equation}
one finds
\begin{equation}  
\xi_{\bbox{k}} = \frac{D}{\sqrt{2}} \sqrt{\sin^2 k_x + \cos^2 k_y},
\label{spec}
\end{equation}
where $D=4\sqrt{2}J \Delta$ is the half-width of the band
and the lattice constant has been set equal to unity. 
At low energies the density of states is determined by
the neighborhood of the two isolated roots of Eq.\ (\ref{spec}) and
exhibits a V-shaped pseudogap $\rho^0(\omega) = 
2 |\omega|/(\pi D^2)$
corresponding to the one found in the flux phase of conventional
fermion mean-field theories. \cite{AFF88,KOT88}

The bandwidth parameter $D$ in Eq.\ (\ref{spec}) is obtained from 
the self-consistency relation
\begin{equation}
\label{sc}
D = \frac{2\sqrt{2}J}{N} \sum_{\bbox{k}} \eta_{\bbox{k}}
\tanh\left(\frac{D \eta_{\bbox{k}}}{2\sqrt{2}T}\right),
\end{equation}
where $\eta_{\bbox{k}} = (\sin^2 k_x + \cos^2 k_y)^{1/2}$,
$N$ is the number of sites, and $k_B = 1$. 
In the limit of $T\rightarrow 0$ Eq.\ (\ref{sc}) reduces to
\begin{equation}
\label{bw0}
D_0 = \frac{2\sqrt{2}J}{N} \sum_{\bbox{k}} \eta_{\bbox{k}},
\end{equation}
yielding a numerical value of $D_0 \approx 1.355 \, J$. The 
correction $\delta D(T)$ to $D_0$ for small but finite temperatures is 
given by
\begin{equation}
\label{bw1}
\delta D(T) = -\frac{32 J}{\pi} \left(\frac{T}{D_0}\right)^3.
\end{equation}

In the following we study the effect of a spin vacancy
located at site 0 (sublattice {\it A})
on the local magnetic properties of the system described by 
Hamiltonian (\ref{Dia}). We simulate this spin defect by decoupling 
site 0 from the rest of the system.
This is done in two steps: First, the 
drone-fermion bond parameter connecting the impurity
site with its nearest neighbors is set to zero,
$i\Delta_{0,\bbox{\delta}} = 
\langle\chi_{\bbox{\delta}} \chi_0\rangle = 0$. This
decouples the drone fermions of site 0
from the rest of the system. Then, 
a local chemical potential $\lambda_0 \rightarrow \infty$ 
acting on site 0 is added to Hamiltonian (\ref{Dia}),
\begin{equation}
H_{\mbox{\scriptsize imp}} = H_{\mbox{\scriptsize MF}} + 
\lambda_0 f_0^{\dagger} f_0.
\label{IMP}
\end{equation}
It induces an empty site at $\bbox{R} = 0$
with respect to the spinon $f$ operators and suppresses the 
low-energy $f$-fermion degrees of freedom at this site.
We will show below that the presence of the local potential also
leads to a vanishing spinon bond parameter connecting the
impurity site with its nearest neighbors, 
$i\tilde{\Delta}_{0,\bbox{\delta}} =
\langle f_{\bbox{\delta}}^{\dagger} f_0\rangle=0$, and therefore
completely decouples the impurity.
Further, one infers $\tilde{\Delta}_{\bbox{R},\bbox{R}+\bbox{\delta}}
= \Delta_{\bbox{R},\bbox{R}+\bbox{\delta}}$ for all 
$\bbox{R}\in A$ which manifests the equivalent treatment
of $f$ and drone fermions as required by spin rotational symmetry.

\section{Local Moment}
%
%
In this section we study the uniform properties and the 
spatial distribution of a local moment induced by a spin vacancy.
We first calculate the impurity contributions to the spinon density of 
states and the uniform spin susceptibility
by means of the uniform Green's function $G(i\omega_n) = 
-1/N\sum_{\bbox{r}} \langle T_{\tau} f_{\bbox{r}}(\tau) 
f^{\dagger}_{\bbox{r}}(0)\rangle_{\omega_n}$ given by
\begin{equation}
G(i\omega_n) = N G^0(i\omega_n) + \delta G(i\omega_n)
\end{equation}
with fermionic frequencies $\omega_n=(2n+1)\pi T$.
The Green's function in the absence of the impurity
$G^0(i\omega_n)$ and the correction $\delta G(i\omega_n)$ due to the
impurity are given by 
\begin{eqnarray}
G^0(i\omega_n) &=& \frac{1}{N} \sum_{\bbox{k},\nu} 
g_{\nu}^0(i\omega_n;\bbox{k}), \nonumber \\
\delta G(i\omega_n) &=& \frac{1}{N}
\sum_{\bbox{k},\nu} g_{\nu}^0(i\omega_n;\bbox{k})
T(i\omega_n) g_{\nu}^0(i\omega_n;\bbox{k})
\label{Guni}
\end{eqnarray}
with $g_{\nu}^0(i\omega_n;\bbox{k}) = 1/(i\omega_n-(-1)^{\nu} 
\xi_{\bbox{k}})$,
$\nu \in \{1,2\}$. The $T$ matrix in Eq.\ (\ref{Guni}) describes 
successive scattering at the vacancy and is obtained from usual scattering
theory; in the limit $\lambda_0\rightarrow \infty$ it is
$T(i\omega_n) = -1/G^0(i\omega_n)$, yielding
\begin{equation}
\delta G(i\omega_n) = \frac{\partial}{\partial i\omega_n} \ln
G^0(i\omega_n).
\label{dG}
\end{equation}
Evaluating Eq.\ (\ref{dG}) one obtains
\begin{equation}
\delta G(i\omega_n) = \frac{1}{i\omega_n}\left(1-\ln^{-1}
\frac{D}{|\omega_n|}\right).
\label{dG2}
\end{equation}

From Eq.\ (\ref{dG2}) we calculate the impurity correction to the 
density of states $\delta\rho(\omega) = -1/\pi \, \mbox{Im} \, 
\delta G(\omega+i0^+)$:
\begin{equation}
\delta \rho(\omega) = \delta(\omega) - \frac{1}{2|\omega|} \,
\frac{1}{\left(\pi/2\right)^2+\ln^2 \left(D/|\omega|\right)}.
\label{DOSu}
\end{equation}
The first and second terms on the right-hand side of Eq.\ (\ref{DOSu}) 
describe, respectively, the formation of a spinon bound state at the 
Fermi surface and the destruction of a spinon singlet in the 
spin background.
Integrating Eq.\ (\ref{DOSu}) over all occupied states one 
finds a reduction in the number of particles by $1/2$. This is in 
agreement with the fact that the local chemical 
potential $\lambda_0$ induces an empty site at $\bbox{R} = 0$
with respect to the $f$ fermions, while in the absence of $\lambda_0$ 
the average $f$-fermion occupation number is $1/2$.
 
With the help of Eq.\ (\ref{dG2}) we evaluate the impurity 
contribution $\delta \chi(T)$
to the uniform spin susceptibility by employing the relation 
\begin{equation}
\delta \chi(T) = T\sum_{i\omega_n} 
\frac{\partial}{\partial i \omega_n} \delta G(i\omega_n).
\label{UNS}
\end{equation}
For $T \ll J$ one obtains 
\begin{equation}
\delta\chi(T) = \frac{1}{4T} \left(
1 -  \ln^{-1} \frac{D}{T} \right).
\label{SUSu}
\end{equation}
The formation of a spinon bound state is thus found to lead to 
a Curie-like spin susceptibility of a free spin $1/2$ with 
logarithmic correction. We note that conventional fermion
mean-field theory \cite{KRI95b}
only reproduces one half of the above result. This
shortcoming has its origin in the mean-field treatment of the local 
constraints on the pseudoparticle occupation number which is 
characteristic for those theories.

We analyze the spatial distribution of the impurity-induced moment  
by means of the local Green's function 
$G(i\omega_n;\bbox{R},\bbox{R}) = -\langle T_{\tau} f_{\bbox{R}}(\tau)
f^{\dagger}_{\bbox{R}}(0)\rangle_{\omega_n}$ which is given by
\begin{equation}
G(i\omega_n;\bbox{R},\bbox{R}) = G^0(i\omega_n) + 
\delta G(i\omega_n;\bbox{R}, 
\bbox{R}).
\label{Gaa}
\end{equation}
Depending upon whether $\bbox{R}$ lies on the {\it A} or {\it B} sublattice
the impurity contribution $\delta G(i\omega_n;\bbox{R},\bbox{R})$ is
\begin{eqnarray}
\delta G(i\omega_n,\bbox{R},\bbox{R}) &=& |A(i\omega_n;\bbox{R})|^2 /
G^0(i\omega_n),
\quad \bbox{R} \in A, \nonumber\\
\delta G(i\omega_n,\bbox{R},\bbox{R}) &=& -|B(i\omega_n;\bbox{R})|^2 /
G^0(i\omega_n),
\quad \bbox{R} \in B
\label{GR}
\end{eqnarray}
with 
\begin{eqnarray}
A(i\omega_n;\bbox{R}\in A) &=& \frac{1}{N} \sum_{\bbox{k},\nu}
g_{\nu}^0(i\omega_n;\bbox{k}) \exp(i\bbox{k}\bbox{R}),\nonumber\\
B(i\omega_n;\bbox{R}\in B) &=& \frac{1}{N} \sum_{\bbox{k},\nu}
g_{\nu}^0(i\omega_n;\bbox{k}) (-1)^{\nu+1} \nonumber\\
&&\mbox{}\times\exp(-i\varphi_{\bbox{k}}+i\bbox{k}\bbox{R}).
\label{AB}
\end{eqnarray}
Evaluating Eqs.\ (\ref{GR}) one finds
\begin{eqnarray}
\delta G(i\omega_n;\bbox{R},\bbox{R}) &=& 
\frac{4\Phi_A(\bbox{R})}{\pi D^2}
K_0^2\left(\frac{\sqrt{2} R |\omega_n|}{D}\right)\nonumber\\
&&\mbox{}\times i\omega_n \ln^{-1} \frac{D}{|\omega_n|} 
\quad \mbox{for} \, \bbox{R} \in A, \nonumber\\
\delta G(i\omega_n;\bbox{R},\bbox{R}) &=&
-\frac{4\Phi_B(\bbox{R})}{\pi D^2}
K_1^2\left(\frac{\sqrt{2} R |\omega_n|}{D}\right)\nonumber\\
&&\mbox{}\times i\omega_n \ln^{-1} \frac{D}{|\omega_n|} 
\quad \mbox{for} \, \bbox{R} \in B,
\label{GRev}
\end{eqnarray}
where the $K_i$ are MacDonald's functions \cite{GRA94}
and the angular
dependence is determined by
\begin{eqnarray}
\Phi_A(\bbox{R}\in A) &=& \left(1+\cos \pi R_x\right)/2,\nonumber\\
\Phi_B(\bbox{R}\in B) &=& \left(1- (2R_x^2/R^2-1) \cos \pi R_x\right)/2.
\end{eqnarray}

From Eqs.\ (\ref{GRev}) we determine the  
impurity contribution to the local density of states, 
$\delta\rho(\omega; \bbox{R}) = -1/\pi \, \mbox{Im} \,
\delta G(\omega + i0^+;\bbox{R},\bbox{R})$.
It is found that the local moment resides predominantly on the
{\it B} sublattice; for $R|\omega|\ll J$ one obtains
\begin{eqnarray}
\delta\rho(\omega;\bbox{R}\in B) &=& 
\frac{2\Phi_B(\bbox{R})}
{\pi R^2} \Big( \frac{\delta(\omega)}{\ln\left(D/|\omega|\right)}
+ \frac{1}{2|\omega|} \nonumber\\
&&\times\frac{1}{\left(\pi/2\right)^2 + \ln^2 
\left(D/|\omega|\right)} \Big),
\label{DOSl}
\end{eqnarray}
while $\delta \rho(\omega;\bbox{R}\in A)$ is negligible.

Our particular interest is the local spin polarization in the
vicinity of the impurity as this quantity is needed below
to derive an expression for the NMR Knight shift which we will
compare to experimental data. In the presence of a uniform magnetic
field the spin polarization on site $\bbox{R}$
can be obtained from the local spin 
susceptibility 
\begin{equation}
\delta \chi(T;\bbox{R}) = T \sum_{i\omega_n} 
\frac{\partial}{\partial i\omega_n} 
\delta G(i\omega_n;\bbox{R},\bbox{R}).
\label{LOS}
\end{equation}
Evaluating Eq.\ (\ref{LOS}) for $R|\omega|\ll J$ 
with $\bbox{R} \in B$ one obtains
\begin{equation}
\delta \chi(T;\bbox{R}\in B) =
\frac{1}{2\pi T} \, \frac{\Phi_B(\bbox{R})}{R^2}
f(T,R)  \ln^{-1}\frac{D}{T}
\label{SUSl}
\end{equation}
with a correction factor 
\begin{equation}
f(T,R) = \left(xK_1(x)\right)^2, \quad 
x = \frac{4 \sqrt{2} T R}{\pi D},
\end{equation}
which approaches unity for $T R\ll J$.
Due to the $R^{-2}$ decay of $\delta\chi(T;\bbox{R})$ the 
spin polarizability is found to be strongest on the 
nearest-neighbor sites of the vacancy. 
Contributions from sublattice {\it A} are again negligible.
The logarithmic correction in Eq.\ (\ref{SUSl})
is due to the marginal character of the bound state:
its spectral weight at a given lattice site vanishes
logarithmically as $\omega\rightarrow 0$. Integrating Eq. (\ref{SUSl})
over $\bbox{R}$ one again obtains the contribution to the uniform 
susceptibility $\delta \chi(T)=1/(4T)$ as required.

We now turn to the 
spinon bond parameter $i\tilde{\Delta}_{\bbox{R},\bbox{R}+
\bbox{\delta}} = \langle f_{\bbox{R}+\bbox{\delta}}^{\dagger} 
f_{\bbox{R}}\rangle$ 
defined for $\bbox{R} \in A$,
which we express in terms of the Green's function
$G(i\omega_n;\bbox{R},\bbox{R}+\bbox{\delta}) = -
\langle T_{\tau} f_{\bbox{R}}(\tau) 
f_{\bbox{R}+\bbox{\delta}}^{\dagger}(0)\rangle_{\omega_n}$:
\begin{equation}
\tilde{\Delta}_{\bbox{R},\bbox{R}+\bbox{\delta}} = 
-i T \sum_{i\omega_n} G(i\omega_n;\bbox{R},\bbox{R}+\bbox{\delta}).
\end{equation}
Splitting the Green's function into an unperturbed
and an impurity-correction term one obtains
\begin{equation}
\tilde{\Delta}_{\bbox{R},\bbox{R}+\bbox{\delta}} =
\tilde{\Delta}_{\bbox{\delta}} + 
\delta\tilde{\Delta}_{\bbox{R},\bbox{R}+\bbox{\delta}}
\end{equation}
with
\begin{eqnarray}
\tilde{\Delta}_{\bbox{\delta}} &=& 
-i T \sum_{i\omega_n} G^0(i\omega_n;0,\bbox{\delta}),
\nonumber\\
\delta\tilde{\Delta}_{\bbox{R},\bbox{R}+\bbox{\delta}} &=&
-i T \sum_{i\omega_n} \delta G(i\omega_n;\bbox{R},\bbox{R}+\bbox{\delta}),
\end{eqnarray}
where
\begin{eqnarray}
G^0(i\omega_n;0,\bbox{\delta}) &=& B(i\omega_n;\bbox{\delta}),
\nonumber\\
\delta G(i\omega_n;\bbox{R},\bbox{R}+\bbox{\delta}) &=&
A^*(i\omega_n;\bbox{R})\nonumber\\
&&\mbox{}\times
B(i\omega_n;\bbox{R}+\bbox{\delta})/ G^0(i\omega_n).
\nonumber\\
\end{eqnarray}
For $\bbox{R}=0$ one finds 
$G^0(i\omega_n;0,\bbox{\delta}) = -\delta G(i\omega_n;0,\bbox{\delta})$
which reconfirms the vanishing spinon bond parameter connecting
the impurity site with its nearest neighbors, i.e., 
$\tilde{\Delta}_{0,\bbox{\delta}} = 0$.

\section{Knight Shift}
%
%
Experimentally, the spatial distribution of an impurity induced moment 
can be investigated by means of the NMR Knight shift $K(T)$ 
on either the impurity, Cu, or in the case of YBa$_2$Cu$_3$O$_7$ 
on the Y nuclei. 
Further information is obtained from the width of the NMR lines
and the relaxation rates of the nuclear magnetic moment.

The Hamiltonian describing a nuclear spin $\bbox{I}$ in an external 
magnetic field $\bbox{H}_0$ coupled to the surrounding electron spins
$\bbox{S}_i$ is 
\begin{equation}
H_N = \gamma_N \bbox{H}_0 
\cdot \bbox{I} + 
A_{\mbox{\scriptsize hf}} \sum_{i} \bbox{S}_{i} \cdot \bbox{I},
\label{NUC}
\end{equation}
where $\gamma_N$ 
denotes the nuclear gyromagnetic ratio and
$A_{\mbox{\scriptsize hf}}$ is the coupling constant of the isotropic
(supertransferred) hyperfine interaction.
In the presence of a nonmagnetic impurity 
the local spin susceptibility that determines the on-site
spin polarization is altered according to Eq.\ (\ref{SUSl}),
yielding for the impurity contribution to the Knight shift
at distance $\bbox{R}$ from the vacancy
\begin{equation}
\delta K(T;\bbox{R}) = \frac{\gamma A_{\mbox{\scriptsize hf}}}
{\gamma_N} 
\sum_{i} \delta\chi(T;\bbox{R}+\bbox{r}_i),
\label{KNIa}
\end{equation}
where the sum is over nearest-neighbor lattice sites of
the nucleus being probed and $\gamma = g \mu_B$
with the $g$ factor of the Cu$^{2+}$ spin and
the Bohr magneton $\mu_B$.
Taking into account only coupling to the most dominant moments 
that reside on the nearest-neighbor sites surrounding the vacancy,
Eq.\ (\ref{KNIa}) becomes
\begin{equation}
\delta K(T) = \frac{\gamma A_{\mbox{\scriptsize hf}}}
{\gamma_N} \, \frac{n}{2 \pi T} 
f(T) \ln^{-1}\frac{D}{T}.
\label{KNI}
\end{equation}
The correction factor
\begin{equation}
f(T) = \left(y K_1(y)\right)^2, \quad 
y = \frac{4 \sqrt{2} T}{\pi D},
\end{equation}
reduces to unity in the limit $T \ll J$.
The variable $n$ in Eq.\ (\ref{KNI}) denotes the number
of sites that are nearest neighbors to the vacancy as well as
to the nucleus being probed.
 
\begin{figure}
\noindent
\centering
\epsfxsize=\linewidth
\epsffile{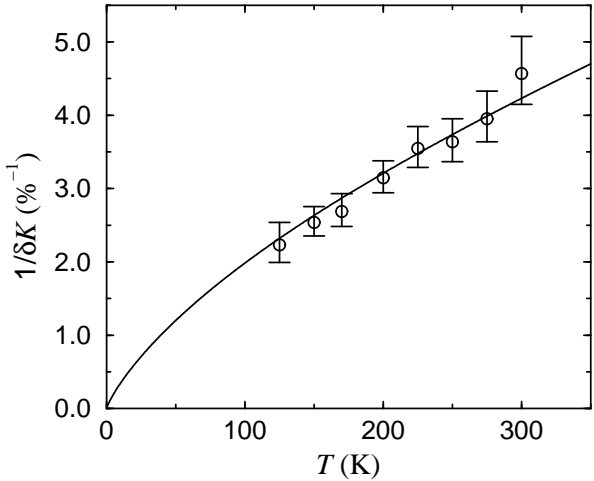}
\caption{Temperature-dependence of the inverse of the 
impurity-induced Knight shift $1/\delta K(T)$ on the 
impurity nucleus. The solid line is a fit employing
Eq.\ (\ref{KNI}) to experimental data by Ishida {\it et al}.\ 
for 3\% $^{27}$Al-doped La$_{1.85}$Sr$_{0.15}$CuO$_{4}$
indicated by circles.}
\label{FIG_Knight}
\end{figure}

The Knight shift on the impurity nucleus is influenced 
predominantly by the
local moments residing on the four nearest-neighbor sites of
the impurity as described by 
Eq.\ (\ref{KNI}) with $n=4$. Figure \ref{FIG_Knight} contains 
the calculated $T$ dependence of $1/\delta K(T)$. We set $J=1500$ K
and $g=2$; by varying 
the supertransferred hyperfine coupling constant 
$A_{\mbox{\scriptsize hf}}$ the curve
is fitted to experimental data obtained by Ishida {\it et al}.\ \cite{ISH96}
for the Knight shift of $^{27}$Al performed on 3\% $^{27}$Al-doped
La$_{1.85}$Sr$_{0.15}$CuO$_{4}$. 
From the fit we find for the supertransferred hyperfine coupling 
constant a numerical value of $^{27}A_{\mbox{\scriptsize hf}}/
\gamma_{\mbox{scriptsize N}} = 
18.2$ k$\,$Gs.
This compares well with $^{27}A_{\mbox{\scriptsize hf}}/
\gamma_N = 16$ k$\,$Gs 
found by Ishida {\it et al}.

Upon doping the system with impurities,
the NMR signal of Cu and Y is observed to split into a main 
line accompanied by satellite line(s) as is discussed in the following.
Within the approximation employed in Eq.\ (\ref{KNI}) we
assume the impurity-induced moments
to reside predominantly on the nearest-neighbor sites of the 
vacancy.  Resonance signals originating from
nuclei that do not lie in the vicinity of the impurity are
affected only little by these moments and contribute to 
the NMR main line. Resonance signals from Cu nuclei on 
nearest-neighbor sites of the impurity are shifted 
with respect to the main line as described by Eq.\ (\ref{KNI})
with $n=1$ and constitute a single Cu satellite line. Resonance 
signals from Y nuclei located on nearest-neighbor and next-nearest 
neighbor sites of the impurity are seen as two satellite lines 
corresponding to $n=2$ and $n=1$, respectively.
These Y satellite lines have been observed 
experimentally by Mahajan {\it et al}.\ \cite{MAH94} in 1\% Zn-doped 
YBa$_2$Cu$_3$O$_{6.64}$.

We now discuss briefly the NMR line-broadening which results from
the inhomogeneous distribution of local Knight shifts in the presence 
of many vacancies. According to Eq.\ (\ref{SUSl}) moments
induced by a vacancy located on the {\it A} sublattice 
are found on all sites of sublattice {\it B} with magnitude $\propto R^{-2}$, 
$R$ being the distance from the vacancy. As a result the 
nuclear resonance signals contributing to the main and satellite lines 
vary with distance from the impurities which can be observed as a broadening
of the line. At a finite concentration $c$ of vacancies the
linewidth $\Delta H$ can be estimated from Eq.\ (\ref{KNIa}),
assuming an average distance between vacancies of 
$R_{\mbox{\scriptsize av}}=\sqrt{1/c}$:
\begin{equation}
\frac{\Delta H}{H_0} \approx 
\frac{\gamma A_{\mbox{\scriptsize hf}}}
{\gamma_N} \, 
\frac{c}{2\pi T} \ln^{-1}\frac{D}{T}.
\end{equation}

\section{Impurity Interaction}
%
%
In this section we study the interaction between two vacancies,
first by considering two isolated impurities. The case of 
a finite concentration of impurities is treated subsequently.
 
The spinon Green's function in the presence
of two impurities located at sites $0$ and $\bbox{R}$ is
\begin{equation}
G_{\bbox{R}}(i\omega_n) = N G^0(i\omega_n) + 
\delta G_{\bbox{R}}(i\omega_n).
\end{equation}
Employing scattering theory the two-impurity correction 
$\delta G_{\bbox{R}}(i\omega_n)$ can be expressed as
\begin{eqnarray}
\lefteqn{\delta G_{\bbox{R}}(i\omega_n) = }\nonumber\\
&&\frac{1}{N} \sum_{\bbox{r}}
\left(
\begin{array}{c}
G^0(i\omega_n;\bbox{r},0)\\
G^0(i\omega_n;\bbox{r},\bbox{R})
\end{array}
\right)
\bbox{T}_{\bbox{R}}(i\omega_n)
\left(
\begin{array}{c}
G^0(i\omega_n;0,\bbox{r})\\
G^0(i\omega_n;\bbox{R},\bbox{r})
\end{array}
\right)
\nonumber\\
\label{GSC}
\end{eqnarray}
with the unperturbed Green's function 
$G^0(i\omega_n;\bbox{R},\bbox{R}') = -\langle T_{\tau} f_{\bbox{R}}(\tau)
f^{\dagger}_{\bbox{R}'}(0)\rangle^0_{\omega_n}$ and the scattering matrix
\begin{equation}
\bbox{T}_{\bbox{R}}(i\omega_n) =
\left(
\begin{array}{cc}
T_{11} & T_{12} \\
T_{21} & T_{22} 
\end{array}
\right)
\end{equation}
with
\begin{eqnarray}
T_{11} &=& T_{22} = -1/G(i\omega_n,\bbox{R},\bbox{R}),\nonumber\\
T_{12} &=& G^0(i\omega_n;0,\bbox{R})/
\left[G(i\omega_n;\bbox{R},\bbox{R}) G^0(i\omega_n)\right], \nonumber\\
T_{21} &=& G^0(i\omega_n;\bbox{R},0)/ 
\left[G(i\omega_n;\bbox{R},\bbox{R}) G^0(i\omega_n)\right].
\label{GT2}
\end{eqnarray}
The diagonal and off-diagonal elements of $\bbox{T}_{\bbox{R}}(i\omega_n)$
describe successive scattering by one impurity and by the two 
impurities, respectively.
From Eqs.\ (\ref{GSC})-(\ref{GT2}) one obtains
\begin{equation}
\delta G_{\bbox{R}}(i\omega_n) =
\frac{\partial}{\partial i\omega_n} \ln \left(
G^0(i\omega_n) G(i\omega_n;\bbox{R},\bbox{R})\right),
\label{G2}
\end{equation}
where $G(i\omega_n;\bbox{R},\bbox{R})$ is given by Eq.\ (\ref{Gaa}).

By means of Eq.\ (\ref{G2}) 
we evaluate the two-impurity contribution to the density of states 
$\delta\rho_{\bbox{R}}(\omega) = -1/\pi \, \mbox{Im} \,
\delta G_{\bbox{R}}(\omega+0^+)$.
For $\bbox{R} \in B$ one finds a splitting of the resonance energy level 
\begin{equation}
\delta \rho_{\bbox{R}\in B}(\omega) = 
\delta\left(\omega-J(\bbox{R})\right) 
+ \delta\left(\omega+J(\bbox{R})\right),
\label{SPL}
\end{equation}
where the parameter $J(\bbox{R})$ controlling the
level splitting is given by
\begin{equation}
J(\bbox{R}) = \frac{D \sqrt{2\Phi_B(\bbox{R})}}{2R}
\ln^{-1} \frac{\pi R}{\sqrt{2\Phi_B(\bbox{R})}}.
\label{ATR}
\end{equation}  
The level splitting results in an attraction between the two impurities
due to the formation of a singlet of the induced moments.
For the shortest distance between vacancies, $R=1$, Eq.\ (\ref{ATR})
gives $J(\bbox{R}) \approx J$ as expected: $J$ is the only 
energy scale of the model under consideration. In the 
present mean-field theory no attraction is found between impurities
lying on different sublattices, i.e., $\bbox{R} \in A$.
These findings are also consistent with exact diagonalization
data of Bulut {\it et al}.\ \cite{BUL89} who found a binding energy
of $-0.56 \, J$ for nearest-neighbor static vacancies on a 
$4 \times 4$ lattice, while the interaction between vacancies
separated by $\bbox{R}$ is negligible if $R_x+R_y$ is an even number.
We interprete $J(\bbox{R})$ given by Eq.\ (\ref{ATR}) as being
the antiferromagnetic exchange coupling of impurity moments.
Indeed, the impurity contribution to the spin susceptibility 
calculated from Eq.\ (\ref{G2}) is
\begin{eqnarray}
\delta\chi(T\gg J(\bbox{R})) &=& \frac{1}{2T},\nonumber\\
\delta\chi(T\ll J(\bbox{R})) &=& \frac{2}{T} 
\exp\left(-\frac{J(\bbox{R})}{T}\right),
\end{eqnarray}
which is precisely that of two 1/2 spins coupled by an interaction
$H_{\mbox{\scriptsize int}} = J(\bbox{R}) \bbox{S}_1 \cdot \bbox{S}_2$.

From Eq.\ (\ref{ATR}) the interaction potential between two 
moments is found to fall off slowly as $R^{-1}$ with distance
between the vacancies.
In a real system with finite impurity concentration $c$, however,
one expects this interaction to be screened at large distance
by the presence of other impurities.
To account for this effect we introduce a finite self-energy 
$\Sigma(\omega)$ determined by means of a coherent-potential
approximation (CPA):
\begin{equation}
\Sigma(\omega) G^0(\omega-\Sigma(\omega))+c=0.
\end{equation}
Neglecting the frequency dependence of $\Sigma(\omega)$ we 
approximate the self-energy by 
$\Sigma(\omega)\approx \Sigma(\omega\rightarrow 0) = -i |\Sigma_0''|$,
where $\Sigma_0''$ fulfills
\begin{equation}
-\frac{4}{\pi D^2} \Sigma_0''^2 \ln\frac{D}{|\Sigma_0''|} + c = 0.
\end{equation}

Reevaluating the exchange coupling parameter $J(\bbox{R})$ 
one finds an exponential cutoff in the two-impurity interaction 
potential
\begin{equation}
J(\bbox{R}) \propto \exp\left(-R/R_{\mbox{\scriptsize av}}\right),
\end{equation}
where $R_{\mbox{\scriptsize av}} = \sqrt{1/c}$ is the average distance 
between impurities. For small distances 
$R \ll J$ one recovers again Eq.\ (\ref{ATR}). The present CPA treatment 
gives a qualitative description of the screening of the interaction
potential of two distant impurities, but it neglects important effects 
such as the formation of spin singlets among closely spaced moments;
a complete analysis of the many-impurity problem should
take this effect into account.

Having the interaction scale between moments we can give an
estimation of the relaxation rate $1/T_1$ of nuclear magnetization
resulting from fluctuations of the impurity-induced
moments in the vicinity of the vacancy. 
To be specific let us consider the nuclear-spin relaxation rate $1/T_1$
at the impurity site (obtained, for instance, from the $^{27}$Al-NMR 
signal in an experiment by Ishida {\it et al}. \cite{ISH96}) 
With the supertransferred hyperfine interaction of 
Eq.\ (\ref{NUC}) the nuclear relaxation rate is
\begin{equation}
\frac{1}{T_1} \approx \frac{2}{3} \tau A^2_{\mbox{\scriptsize hf}}  
\sum_{\bbox{\delta}\bbox{\delta}'} \langle \bbox{S}_{\bbox{\delta}}
\cdot \bbox{S}_{\bbox{\delta}'}\rangle,
\label{REL}
\end{equation}
where $\tau$ is the correlation time of the local moment and 
$\hbar = 1$. The amplitude of spin fluctuations on 
nearest-neighbor sites of the impurity is given by the equal-time 
spin-correlation function 
\begin{eqnarray}
\lefteqn{\sum_{\bbox{\delta}\bbox{\delta}'} \langle 
\bbox{S}_{\bbox{\delta}} \cdot \bbox{S}_{\bbox{\delta}'}\rangle}\nonumber\\
&=& -3 T^2 \sum_{i\omega_n, i\nu_m}\sum_{\bbox{\delta}\bbox{\delta}'}
\delta G(i\omega_n+i\nu_m;\bbox{\delta},\bbox{\delta}')
\delta G(i\omega_n;\bbox{\delta}',\bbox{\delta}) \nonumber\\
&=& 3\left(\frac{z}{\pi}\right)^2
\ln^{-2}\frac{D}{T},
\end{eqnarray}
where $\delta G(i\omega_n;\bbox{\delta},\bbox{\delta}') =
-B^*(i\omega_n;\bbox{\delta}) B(i\omega_n;\bbox{\delta}') /
G^0(i\omega_n)$, $\nu_m = 2m\pi T$ are bosonic frequencies, and
$z=4$ is the number of nearest neighbors.
We assume the correlation time $\tau$ in Eq.\ (\ref{REL})
to be dominated by the exchange interaction 
among moments. In principle, $\tau$ has a certain distribution due 
to the random spreading of impurities. We do not go into this 
delicate issue and roughly approximate 
$\tau \approx 1/\left[z J(\sqrt{2} R_{\mbox{\scriptsize av}})\right]$,
where $J(\sqrt{2} R_{\mbox{\scriptsize av}})$ is given by Eq.\ 
(\ref{ATR}) and $\sqrt{2} R_{\mbox{\scriptsize av}} = \sqrt{2/c}$ 
is the average distance between impurities located on different 
sublattices. One then obtains
\begin{equation}
\tau \approx \frac{\sqrt{2}}{zD\sqrt{c}} \ln\frac{2\pi^2}{c}.
\label{TAU}
\end{equation}
Assuming an impurity concentration $c=3\%$, a temperature 
$T = 100$ K, and using the value of the supertransferred hyperfine
coupling constant $A_{\mbox{\scriptsize hf}}$ found in the previous section, 
Eqs.\ (\ref{REL})-(\ref{TAU}) give $1/T_1 \approx 0.29$ m$\,$s$^{-1}$. 
We compare this value to $1/T_1 = 0.43$ m$\,$s$^{-1}$
observed by Ishida {\it et al}.\ \cite{ISH96} for 3\% $^{27}$Al-doped
La$_{1.85}$Sr$_{0.15}$CuO$_{4}$ and find it to be of the same order
of magnitude.

\section{Kondo Screening of Local Moments}
%
%
The two-dimensional Heisenberg model has a N\'eel-ordered
ground state and it was implicitly assumed above that the spin-gap disordered
state is stabilized by mobile holes. Naturally, the question then arises
how a local moment induced by a static vacancy will be affected by
these charge carriers. Once the metallic state with finite
Fermi surface is formed one usually expects Kondo screening of the local
moment at low temperature. As an exception, a spinon bound state on a vacancy
can coexist with metallic conductivity if spin-charge separation
occurs with a gap in the spin-excitation spectrum. 
Underdoped cuprates, however, seem
to have ungapped pieces of their original (large) Fermi surfaces even 
in a ``spin-gap'' phase \cite{LOE96}. Therefore the bound spinon on the
vacancy is screened by gapless Fermi-surface excitations, although
the Kondo temperature is expected to be low due to the reduced 
density of states. The characteristic Kondo-energy scale can be roughly
estimated in the following way.

We introduce fermionic field operators $\psi_{i\sigma} = s_{\sigma}
\varphi(\bbox{R}_i)$ which annihilate a spinon $s_{\sigma}$ in a bound
state with wave function $\varphi(\bbox{R}_i)$, 
$\bbox{R}_i$ being the distance
from the spinless vacancy. Due to the hopping term in the {\it t}-{\it J} 
model there is a finite amplitude $t \varphi(\bbox{R}_i)$ of mixing of the 
bound spinon with extended states:
\begin{equation}
H = t\sum_{i,\bbox{\delta},\sigma}\left( 
c^{\dagger}_{i+\bbox{\delta},\sigma} h_i^{\dagger} \psi_{i,\sigma}
+ \mbox{H.c.} \right).
\label{KON1}
\end{equation}
The bosonic operators $h_i^{\dagger}$ and 
$c^{\dagger}_{i+\bbox{\delta},\sigma}$ in Eq.\ (\ref{KON1})
create a ``holon'' and a conduction electron, respectively. In a slave-boson
formulation, the latter is a product of the lattice spinon and holon
operators.

At finite doping, $x = \langle |h_i|^2\rangle\neq 0$, 
the Hamiltonian (\ref{KON1})
allows for a localized spinon to escape the impurity which 
leads to a broadening of the bound state. The width of this resonance
controls the low-temperature behavior of the local spin and is given by
the $s_{\sigma}$ fermion self-energy
\begin{eqnarray}
\Sigma^s(\omega) &=& t^2 T \sum_{\nu} \sum_{ij,\bbox{\delta}
\bbox{\delta}'}
\varphi^*(\bbox{R}_i) \varphi(\bbox{R}_j)
G^h_{\nu}(\bbox{R}_{ij}) \nonumber\\
&&\mbox{}\times G^c_{\omega-\nu}(\bbox{R}_{i+
\bbox{\delta}}-\bbox{R}_{j+\bbox{\delta}'}).
\label{RES}
\end{eqnarray}
At low temperatures the holons are almost condensed, and the holon
Green's function can be approximated by
$G^h_{\nu} \approx (x/T) \delta_{\nu,0}$. Furthermore,
we keep only the nonoscillatory on-site terms that present 
the leading contributions to Eq.\ (\ref{RES})
and employ the normalization condition $\sum_i |\varphi_i|^2 = 1$.
Then the width of the Kondo resonance is
\begin{equation}
\mbox{Im} \, \Sigma^s(0) \approx \pi z x t^2 N^c(0),
\label{KON2}
\end{equation}
where $N^c(0) = 1/\pi \sum_k \mbox{Im}\, G^c_k(\omega=0)$ is the on-site
density of electronic states on the Fermi level. 

In the overdoped {\it t}-{\it J} model with a large gapless Fermi surface  the density 
of states is controlled by the inverse of the full bandwidth, $N^c(0) 
\approx 1/(2zt)$. In the underdoped pseudogap 
regime, $x t \ll J$, the density of states
$N^c(0)$ is strongly reduced which is due to the fact that the ungapped areas
of the Fermi surface contributing to $N^c(0)$ are only small. We simply
assume $N^c(0) \approx x/(2zt)$ and estimate the Kondo temperature 
using Eq.\ (\ref{KON2}) as
\begin{equation}
T_K(x) \approx \frac{\pi}{2} x^2 t.
\label{TK}
\end{equation}

For a typical value of $x \approx 0.1$ and $t \approx 0.4$ eV this 
gives $T_K \approx 70$ K.
$T_K(x)$ quickly increases with $x$, causing the vacancy-induced moment to 
disappear in the optimal and overdoped regime. This 
magnetic-nonmagnetic Kondo crossover upon hole doping was recently 
employed by Nagaosa and Lee \cite{NAG97} to explain the unusual doping
dependence of the residual resistivity of cuprate superconductors.

One may wonder why the exchange parameter $J$ does not enter
in Eq.\ (\ref{TK}). In fact, the main role of the Heisenberg
term in this quite unusual Kondo-like behavior of a spinless
vacancy is to produce a spin pseudogap hence creating
a local moment. It is the hopping term that is then responsible
for converting the spinon bound state into a resonance of
finite width.

\section{Conclusion}
%
%
In summary, we have studied the localized magnetic states induced by
static spin vacancies in underdoped
high-$T_c$ cuprates. Starting from the two-dimensional Heisenberg
antiferromagnet we employ a drone-fermion mean-field theory describing 
a spin liquid in the spin-gap phase. The impurity induces  
a spinon bound state residing predominantly on the 
nearest-neighbor sites of the impurity.  
We have calculated the impurity-induced Knight shift and found it
in good agreement with 
experimental data of Ishida {\it et al}.\ \cite{ISH96} for Al-doped
cuprates. The calculations do also account for the appearance of 
satellite peaks in NMR measurements on Y observed by Mahajan {\it et al}.\
\cite{MAH94} in Zn-doped cuprates.
Two impurities are found to attract each other due to the
formation of a singlet of the induced moments. In the presence
of a finite concentration of impurities the interaction between 
moments is screened at distances larger then the average separation between
impurities. We estimate the contribution of local moments to the NMR 
relaxation rate which is found to be in reasonable agreement with 
experiment. Finally, an estimation of the Kondo temperature of 
screening of the vacancy-induced moments by charge carriers 
is given, below which a 
nonmagnetic impurity behaves as a sharp resonance at the Fermi level.


\end{document}